\documentclass[aps,prl,reprint,superscriptaddress]{revtex4-1}

\usepackage[utf8]{inputenc}
\usepackage{amsmath,amsfonts,amssymb}
\usepackage{todonotes}
\usepackage{graphicx} \graphicspath{{./Figures/}}
\usepackage{braket}
\usepackage[hypertexnames=false,colorlinks=true,linkcolor=blue,citecolor=blue,
            bookmarksopen=false,bookmarks=false,
            pdfstartview=XYZ,pdffitwindow=true,pdfcenterwindow=true]{hyperref}

\newcommand{\RSet}{\mathbb{R}}

\newcommand{\epsi}{\varepsilon}
\newcommand{\sph}{\mathbb{S}^2}
\newcommand{\vect}[1]{\ensuremath{\bm #1}}

\newcommand{\vx}{\vect{x}}

\newcommand{\vxi}{\vect{\xi}}
\newcommand{\vy}{\vect{y}}
\newcommand{\grad}{\vect{\nabla}}

\newcommand{\edits}[1]{#1}

\begin{document}

\title{Spatio-temporal canards in neural field equations}

\author{D. Avitabile}
\affiliation{Centre for Mathematical Medicine and Biology, School of Mathematical
Sciences, University of Nottingham, University Park, Nottingham NG9 7RD, UK}

\author{M. Desroches}
\affiliation{Inria Sophia Antipolis M{\'e}diterran{\'e}e Research Centre, MathNeuro Team, 2004 route des Lucioles - BP 93\\ 06902 Sophia Antipolis, Cedex, France}

\author{E. Knobloch}
\affiliation{Department of Physics, University of California, Berkeley, CA 94720, USA}

\begin{abstract}
  Canards are special solutions to ordinary differential equations
  that follow invariant repelling slow manifolds for long time intervals. In
  realistic biophysical single cell models, canards are responsible for several
  complex neural rhythms observed experimentally, 
  but their existence and role in spatially-extended systems is largely unexplored. We
  describe a novel type of coherent structure in which a
  spatial pattern displays temporal canard behavior. Using interfacial dynamics
  and geometric singular perturbation theory, we classify spatio-temporal canards
  and give conditions for the existence of folded-saddle and folded-node canards. We
  find that spatio-temporal canards are robust to changes in the synaptic
  connectivity and firing rate. The theory correctly predicts the existence of
  spatio-temporal canards with octahedral symmetry in a neural field model posed
  on the unit sphere.
%
\end{abstract}

\maketitle

\section{\label{sec:intro}Introduction}
Spatially extended, continuum, deterministic neural field models take the
form~\cite{Bressloff2012o,Bressloff:2014cm,Ermentrout:2010cga}
\begin{equation}\label{eq:neuralFieldModel}
  \partial_t u(x,t) = - u(x,t) + \int_{\RSet} W(x,y) f(u(y,t) - h) \, d y,
\end{equation}
where $u$ denotes the coarse-grained activity of a neural population
at position $x\in \RSet$ and time $t \in \RSet^+$, $W$ is a synaptic
kernel modelling the strength of connections from neurons at positions
$y$ to those at position $x$, $f$ is a firing rate function converting
neural activity into synaptic inputs and $h$ is a firing rate
threshold. Nonlocal equations of this type,
originally proposed by Wilson and Cowan~\cite{Wilson1972aa} and
Amari~\cite{Amari1975aa}, provide a coarse-grained model of macroscopic 
brain activity~\cite{Folias2005aa}, and have been used to explain experimental
observations of cortical waves in vitro~\cite{Richardson:2005cs} and in
vivo~\cite{Huang:2004kw,GonzalezRamirez:2015gk}, as well as electroencephalogram
recordings~\cite{SteynRoss:2003ep} and feature selectivity in the primary visual
cortex~\cite{Camperi:1998ji}.

In this article we demonstrate that neural fields described by
Eq.~(\ref{eq:neuralFieldModel}) support generically a novel type of coherent
structure, in which a spatial pattern displays temporal canard
behavior~\cite{Benoit1981}.  We refer to these solutions as \emph{spatio-temporal
canards}. Canards are considered to be a footprint of time scale separation in
ordinary differential equations (ODEs): these special solutions follow
(locally) invariant repelling slow manifolds for long time intervals,
and manifest themselves via $O(1)$ amplitude changes that take place 
within an exponentially small range of parameter values. In planar
systems, this brutal growth of solutions is referred to as a
\emph{canard explosion}~\cite{Benoit1981,Krupa2001}.

It is widely accepted that canards have a \emph{functional} role in
biophysical single-neuron models of Hodgkin--Huxley-type, where they
approximate excitability thresholds~\cite{Desroches2013b,Mitry2013}
and organise abrupt transitions from resting to spiking
states~\cite{Moehlis2006}, or from spiking to bursting
regimes~\cite{Kramer2008,Rinzel1986}. In addition, canards underpin
complex neural rhythms such as mixed-mode
oscillations~\cite{Desroches2012} or spike-adding
phenomena~\cite{Desroches2013} \edits{in bursters}.

\edits{An intriguing open question concerns the existence and role of temporal 
canards in spatially extended dynamical systems with time scale separation. 
Numerical simulations indicate that canards do indeed exist in such systems 
\cite{Gandhi16}, but the absence of a rigorous geometric singular perturbation 
theory near non-hyperbolic slow manifolds for infinite-dimensional dynamical 
systems requires that the interpretation of such computations be treated with 
caution. The reduction procedure described in the first part of this paper 
overcomes this difficulty in a key example.}

\begin{figure*}
  \includegraphics{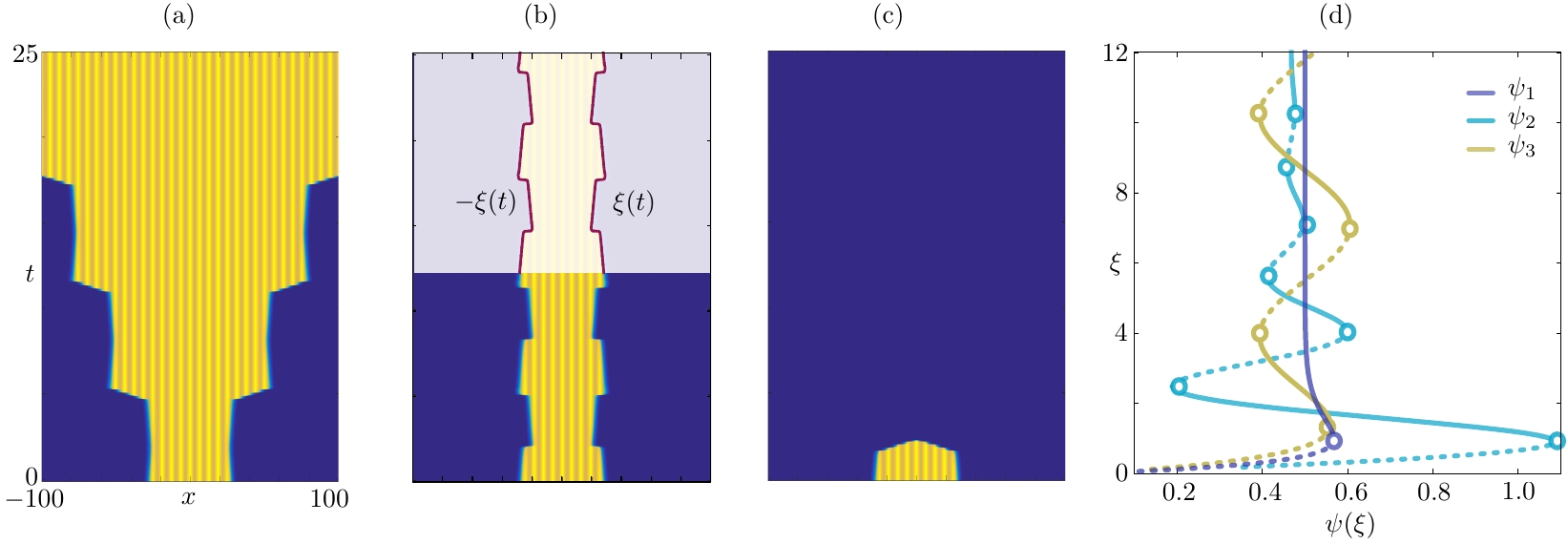}
  \caption{Time simulations of the system~\eqref{eq:extendedNeuralField} for
  $\varepsilon=3.62\!\cdot\!10^{-3}$, $\beta\!=\!\gamma\!=\!0$ and (a) $\alpha\!=0.49$,
  (b) $0.50$, (c) $0.51$; $W$ is as in Eq.~\eqref{eq:heterokernel} with $a=\Lambda=1$
  and $b=0.3$; $\Theta(u)=1/(1+\exp(-50u))$. In (b) we superimpose the threshold crossings
  $x=\pm \xi(t)$ on the pattern, shown for $t\in [12.5,25]$ on a lighter background 
  for better contrast. (d) Examples of the functions
  $\psi_i$ corresponding to kernels $W_i$, $i=1,2,3$, commonly used in neural field
  models: $W_1(x,y)=(1+0.5|x-y|)\exp(-|x-y|)$ is a purely excitatory,
  translation-invariant kernel used, for instance, in Ref.~\cite{Coombes2005aa};
  $W_2(x,y)= \exp(-0.25|x-y|)(0.25 \sin |x-y| + \cos |x-y|)$ is an
  excitatory-inhibitory, oscillatory, translation-invariant kernel used in
  Refs.~\cite{Laing2003aa,Rankin:2014bz}; $W_3$ is the oscillatory heterogeneous
  kernel \eqref{eq:heterokernel} used in (a)--(c) and in all other calculations of
  this paper~\cite{Bressloff:2001ck,Coombes2011aa,Avitabile2015aa}. We plot $\xi$
  on the vertical axis, so that the figure can be read as a bifurcation diagram of
  the full neural field system~\eqref{eq:neuralFieldModel}. Solid (dashed) lines
  indicate stable (unstable) stationary patterns.} 
  \label{fig:figs1-2}
\end{figure*}

In this article we identify canards in neural field models of
the type shown in Eq.~\eqref{eq:neuralFieldModel}. When the threshold 
$h$ is constant, the neural field admits an $h$-dependent family of
coexisting stationary localized solutions organised along a branch
with one or more folds \cite{Laing2002aa,Coombes2003aa}. When $h$ varies slowly
with respect to the macroscopic characteristic time $t$ of
Eq.~\eqref{eq:neuralFieldModel} the system may drift along this branch of equilibria
but abrupt transitions, \edits{excitable} dynamics on the faster time scale $t$, may
occur in the vicinity of the folds where the state of the system `jumps' to a
different state. We remark that the time scales of interest in our study are
different from those used in previous work on canards: these structures have thus far
only been found when there exists a time scale separation at the level of a single cell,
between the membrane potential (fast) and gating variables (slow); in the present
study we find canards in neural fields, which are coarse-grained models of neural
networks, and the time scale separation is between the threshold crossing dynamics
(slow) and the activity variable $u$ (fast).  Our findings can be summarized as
follows: (i) If the firing threshold $h$ varies slowly, complex
spatio-temporal patterns containing canard segments exist for steep
firing rates $f$ and for generic choices of the synaptic kernel $W$;
(ii) A theory for the classification of such spatio-temporal canards
can be derived using interfacial dynamics; (iii) Spatio-temporal
canards of \emph{folded-node} or \emph{folded-saddle} types are
present, depending on the coupling between $h$ and $u$; 
(iv) The behavior described above is robust to changes in the synaptic
kernel $W$ and to perturbations in the firing rate function $f$.

\section{\label{sec:interface}Interface dynamics}
The interfacial description \cite{Coombes2012aa} applies in the case 
$f(u)=\Theta(u)$, where $\Theta(u)$ is the Heaviside step function.
As customary, we consider localized regions of activity 
$-\xi(t)\le x\le\xi(t)$, where the interfaces (or threshold crossings) 
$x=\pm \xi(t)$ are defined by the level set conditions $u(\pm \xi(t),t) = h(t)$ 
with $\partial_x u(\pm \xi(t),t) \lessgtr 0$, for all $t \in \RSet^+$,
and take their width $2 \xi(t)$ as a measure of the spatial
extent of the solution (see for instance Fig.~\ref{fig:figs1-2}(b)). 
Integrating~\eqref{eq:neuralFieldModel}, we find that solutions
$u(x,t)$ can be expressed in terms of the interfacial functions
$\xi(t)$ and the initial datum $u(x,0)$,
\begin{equation}\label{eq:uSolution}
u(x,t) = e^{-t} u(x,0) + \int_0^t \int_{-\xi(s)}^{\xi(s)} e^{s-t} W(x,y) \, dy \, ds.
\end{equation}

\edits{The approach of Refs.~\cite{Coombes2011aa,Coombes2012aa} can be extended to the case of
time-dependent $h$. In this case differentiation of the level set condition for $\xi$
with respect to time leads to a closed scalar evolution equation for the half-width
of the pattern. Using \eqref{eq:neuralFieldModel} we obtain}
\begin{equation}\label{eq:xiEvolution}
\varphi(\xi,t) \dot \xi = h + \dot h - \psi(\xi),
\end{equation}
where $\varphi(\xi,t) = \partial_x u(\xi,t)$ and
$\psi(\xi) = \int_{-\xi}^\xi W(\xi,y) \, dy$. By hypothesis $\varphi$ is 
strictly negative at all times. The function $\psi$ encodes the neural 
connectivity of the model, as it depends solely on the synaptic kernel 
$W$, which models arbitrary heterogeneous synaptic circuits.
Figure~\ref{fig:figs1-2}(d) shows $\psi(\xi)$ for several commonly
used kernels $W(\xi,y)$ and \edits{highlights that $\psi$ generically possesses folds.
These are marked by circles in Fig.~\ref{fig:figs1-2}(d) and correspond to locations
where $\psi'=0$.}
Equation~\eqref{eq:xiEvolution} represents an exact reduction of the field 
equation \eqref{eq:neuralFieldModel} for $u$ with a time-dependent threshold and Heaviside
firing rate, and constitutes a key tool for the study of spatio-temporal
canards. 

If $h$ is a \emph{constant} control parameter, Eq.~\eqref{eq:xiEvolution} 
admits equilibria for all $h$ and $\xi$ such that $h = \psi(\xi)$. In other 
words, the curves in Fig.~\ref{fig:figs1-2}(d) can be interpreted as branches 
of steady (patterned) states of the full system~\eqref{eq:neuralFieldModel} with
the parameter $h$ identified with $\psi(\xi)$ in Fig.~\ref{fig:figs1-2}(d). 
This strategy for constructing patterns, contained in the original work of Amari~\cite{Amari1977aa},
can be extended also to study stability: to each fold of $\psi$ corresponds a
saddle-node bifurcation of the full system. In Ref.~\cite{Avitabile2015aa} it was shown
that \edits{sinusoidal modulation of the kernel in space generates an infinite number of saddle-nodes organized}
in a \emph{snakes-and-ladders} bifurcation structure \cite{Knobloch15}.


The firing rate threshold parameter, $h$, is therefore a common continuation
parameter in neural field studies: as $h$ is varied, we obtain branches of patterned
stationary states and, depending on the choice of the kernel, secondary symmetry-breaking
bifurcations may occur. It is therefore natural to search for canards in cases where
$h$ is slowly varying. Variations of $h$ have been considered before in the
literature: in Ref.~\cite{Brackley2007aa,Thul:2016gr}, the firing threshold was subject to
fluctuations induced by noise, decoupled from the network activity; in
Refs.~\cite{Coombes:2005hp,coombes2007exotic} the threshold $h$ was coupled directly
to the local value of $u$, in order to mimic spike-frequency adaptation, observed
experimentally in \emph{in vitro} experiments of rat pyramidal neurons~\cite{Madison:1984di}.

In the following we study spatio-temporal canards by combining a spatially-extended
neural field with a slowly-varying oscillatory threshold $h(t)$ which may arise, for
instance, from the competition between \emph{adaptation} and \emph{facilitation}
processes, coupled to the neural field via the macroscopic width of the pattern,
and describe a simple example of the dynamics that result when $h$ evolves on a
slow time scale. Depending on the choice of control parameters, we consider limits
where $h$ influences $u$ (but not vice-versa), as well as cases where the dynamics of
$h$ and $u$ are fully coupled, as previously done in
Refs.~\cite{Brackley2007aa,Thul:2016gr} and~\cite{Coombes:2005hp,coombes2007exotic},
respectively.
Specifically, we study the extended neural field model
\begin{equation}\label{eq:extendedNeuralField}
\begin{aligned}
  & \partial_t u(x,t) = - u(x,t) +\!\! \int_{\RSet} W(x,y)\Theta
  \big[u(y,t)\!-\!h(t)\! \big]dy,
  \\
  & \ddot h(t) + \epsi^2 h(t) = \epsi^2 ( \alpha + \beta \xi(t)) + \epsi \gamma \dot \xi(t), \\
  & \xi(t) = \frac{1}{2} \int_\RSet \Theta \big[ u(y,t) - h(t) \big]
  \, dy.
\end{aligned}
\end{equation}
Thus $h$ obeys a weakly forced oscillator equation with a low natural frequency
$\epsi$ that is
coupled to the neural field via both $\xi$ and $\dot \xi$. In
Figs.~\ref{fig:figs1-2}(a)--(c) we show direct simulations of the
model~\eqref{eq:extendedNeuralField}, displaying strong sensitivity to changes in the
parameter $\alpha$. We will show below that spatio-temporal canards organise abrupt
transitions between branches patterned states, and are therefore responsible for the
behavior shown in Figs.~\ref{fig:figs1-2}(a)--(c). 

Interactions between excitable systems and slow oscillations are
known to produce canard-type dynamics in ODEs with folded
saddles~\cite{Mitry2013,desroches2016spike}. This type of interaction motivated our
choice of the coupling in model~\eqref{eq:extendedNeuralField}, which indeed produces
canards in a spatially-extended system.
In terms of the slow time $\tau = \epsi t$, system~\eqref{eq:extendedNeuralField} is
equivalent to
\begin{equation}\label{eq:rescaledPhi}
\begin{aligned}
  \epsi | \varphi_\epsi( \xi, \tau) | \dot \xi & = \psi(\xi) - h - \epsi (q + \gamma \xi), \\
  \dot h & = q + \gamma \xi, \\
  \dot q & = \alpha + \beta \xi - h,
\end{aligned}
\end{equation}
where $\varphi_\epsi$ is a rescaled version of $\varphi$ and we used the fact
that $\varphi$ and $\varphi_\epsi$ are both strictly negative at all times.
Crucially, we passed from model~\eqref{eq:extendedNeuralField}, which involves an
evolution equation for the scalar field $u(x,t)$, to model~\eqref{eq:rescaledPhi}, whose
state variables are the scalars $(\xi(t),h(t))$.
Since $\lim_{\epsi \to 0^+} \epsi | \varphi_\epsi( \xi, \tau) | = 0$ for all $\tau\in \RSet^+$,
Eqs.~\eqref{eq:rescaledPhi} take the form of a singularly perturbed
system, with one fast variable $\xi$ and two slow variables $h$ and $q$. An important
object for understanding the dynamics of such systems is the \textit{critical
manifold} $S^0$, defined as the $\epsi=0$ limit of the fast
nullsurface~\cite{Krupa2001}. In the
present case, this manifold is the folded surface
$\{(h,q,\xi) \in \RSet^3 \colon h = \psi(\xi)\}$. 
The limit yields the differential-algebraic system
\begin{equation}\label{eq:DAE}
\begin{aligned}
         0 & = \psi(\xi) - h,\\
  \dot h & = q + \gamma \xi, \\
  \dot q & = \alpha + \beta \xi - h,
\end{aligned}
\end{equation}
or equivalently the reduced system (or slow subsystem)
\begin{equation}\label{eq:slowFlow}
\begin{aligned}
  -\psi'(\xi) \dot \xi & = - q - \gamma \xi,\\
  \dot q & = \alpha + \beta \xi - \psi(\xi).
\end{aligned}
\end{equation}
This system is singular when $\psi'(\xi) = 0$, that is, at the folds of the critical
manifold separating attracting sheets from repelling ones. For the problem under
consideration, the singularity occurs at fold lines $\{ (h, q, \xi_*) \in \RSet^3
\colon h = \psi(\xi_*),\; \psi'(\xi_*) = 0\}$; in passing we note that $\xi_*$ can be
any of the folds marked by circles in Fig.~\ref{fig:figs1-2}(d).
It is possible to remove this singularity by rescaling time by the factor
$-\psi'(\xi)$, leading to the desingularised reduced system (DRS)
\begin{equation}\label{eq:desingSlowFlow}
\begin{aligned}
  \dot \xi & =  -q - \gamma \xi,\\
  \dot q & = \psi'(\xi) \big[ \psi(\xi) - \alpha -\beta \xi\big].
\end{aligned}
\end{equation}
We carry out this rescaling because it is helpful in deciphering the flow of
system~\eqref{eq:slowFlow} near the fold lines. Indeed, the rescaling has two major
consequences: (i) Orbits of system~\eqref{eq:slowFlow} are extended in
system~\eqref{eq:desingSlowFlow} to the fold lines, where~\eqref{eq:slowFlow} is
undefined; (ii) System~\eqref{eq:slowFlow} may possess equilibria on the fold lines.
As we shall see below, these equilibria are related to canards in
system~\eqref{eq:desingSlowFlow}.

\begin{figure*}
    \includegraphics{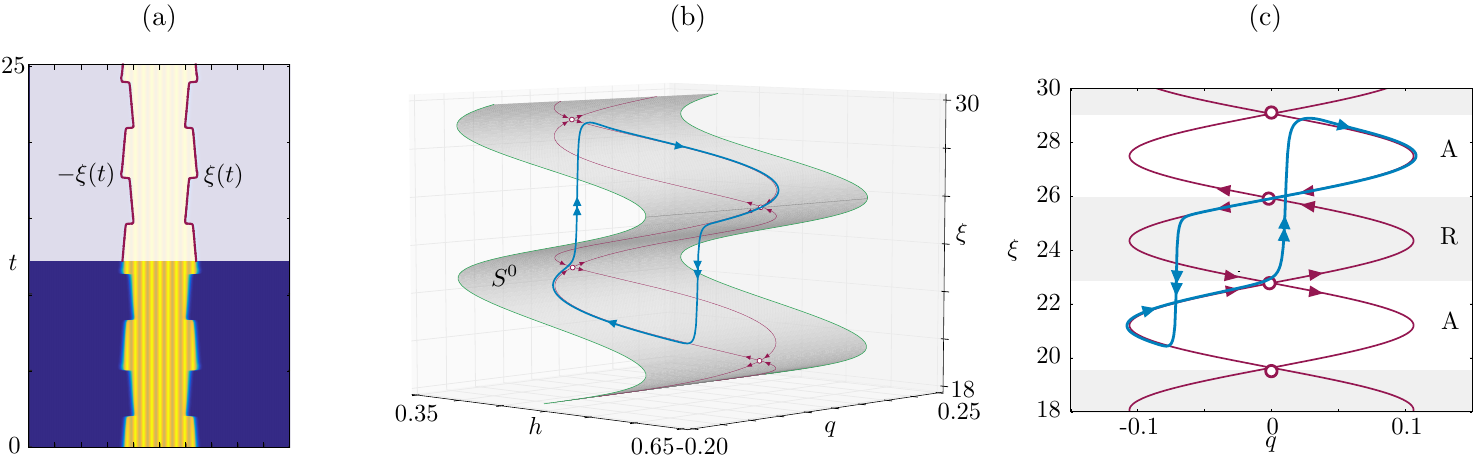}
    \caption{Examples of solutions containing spatio-temporal canards of
    folded-saddle type. (a): Time simulation of the folded-saddle case for
    $\alpha=0.5$, $\beta=\gamma=0$, and remaining parameters as in
    Fig.~\ref{fig:figs1-2}(b). (b): Solution to the full spatio-temporal
    model~\eqref{eq:extendedNeuralField}, also shown in (a) and in
    Fig.~\ref{fig:figs1-2}(b), projected on the \edits{$(h,q,\xi)$} phase space (blue), where
    we also plot the critical manifold $S^0$ (grey) and singular canards
    of~\eqref{eq:slowFlow} on $S^0$ (red). (c): Projection on the \edits{$(q,\xi)$} plane,
    showing the attracting (A) and repelling (R) sheets of $S^0$, revealing a
    spatio-temporal canard.}
    \label{fig:foldedSingCanards}
\end{figure*}

\section{Folded singularities and canards in the extended system} \label{sec:canards}

System~\eqref{eq:desingSlowFlow} has an equilibrium at $(\xi_*,-\gamma \xi_*)$, where
$\xi_*$ satisfies $\psi'(\xi_*) = 0$, i.e., on a fold line of the surface
\edits{$S^0$}. This is not an equilibrium of the reduced system~\eqref{eq:slowFlow}
because of the time rescaling by $-\psi'(\xi)$, which reverses the orientation of
trajectories on the repelling sheets of $S^0$. Therefore solutions to the reduced
system~\eqref{eq:slowFlow} approach the point $(\xi_*,-\gamma \xi_*)$ along an
attracting sheet of $S^0$,
cross it in finite time, and continue to flow along a repelling sheet of $S^0$: these
solutions of system~\eqref{eq:slowFlow} are called \emph{singular canards} and
persist for small $\epsi>0$ as \emph{canard solutions} of
system~\eqref{eq:rescaledPhi}, and hence as \emph{spatio-temporal canards} of
system~\eqref{eq:extendedNeuralField}. 

Equilibria $(\xi_*,-\gamma \xi_*)$ of the DRS~\eqref{eq:desingSlowFlow} are called
\textit{folded singularities} (of node, saddle or focus type) and are therefore
important \edits{in the classification of canards}. Other equilibria of the DRS may exist as true
equilibria of the reduced system~\eqref{eq:slowFlow}: these states are not
generically related to canards and are not considered here. The Jacobian at
$(\xi_*,-\gamma \xi_*)$ is given by
\[
\begin{pmatrix}
  - \gamma & -1  \\
  \pi(\xi_*) & 0
\end{pmatrix},
\qquad
\pi(\xi_*) = \psi''(\xi_*)\big[ \psi(\xi_*) - \alpha -\beta \xi_* \big],
\]
and hence \edits{$(\xi_*,-\gamma \xi_*)$ is either (i) a folded saddle (if $\pi(\xi_*) < 0$)
or (ii) a folded node (if $0 < \pi(\xi_*) < \gamma^2/4$), corresponding in \eqref{eq:rescaledPhi}
to (i) excitable dynamics and (ii) mixed-mode dynamics.}

\edits{Classical theory~\cite{Desroches2012} now guarantees the presence of
canards in~\eqref{eq:rescaledPhi}, and these correspond} to
spatio-temporal canards in~\eqref{eq:extendedNeuralField} for sufficiently
small $\epsi>0$, close to the above-mentioned folded singularities.

\begin{figure*}
  \includegraphics{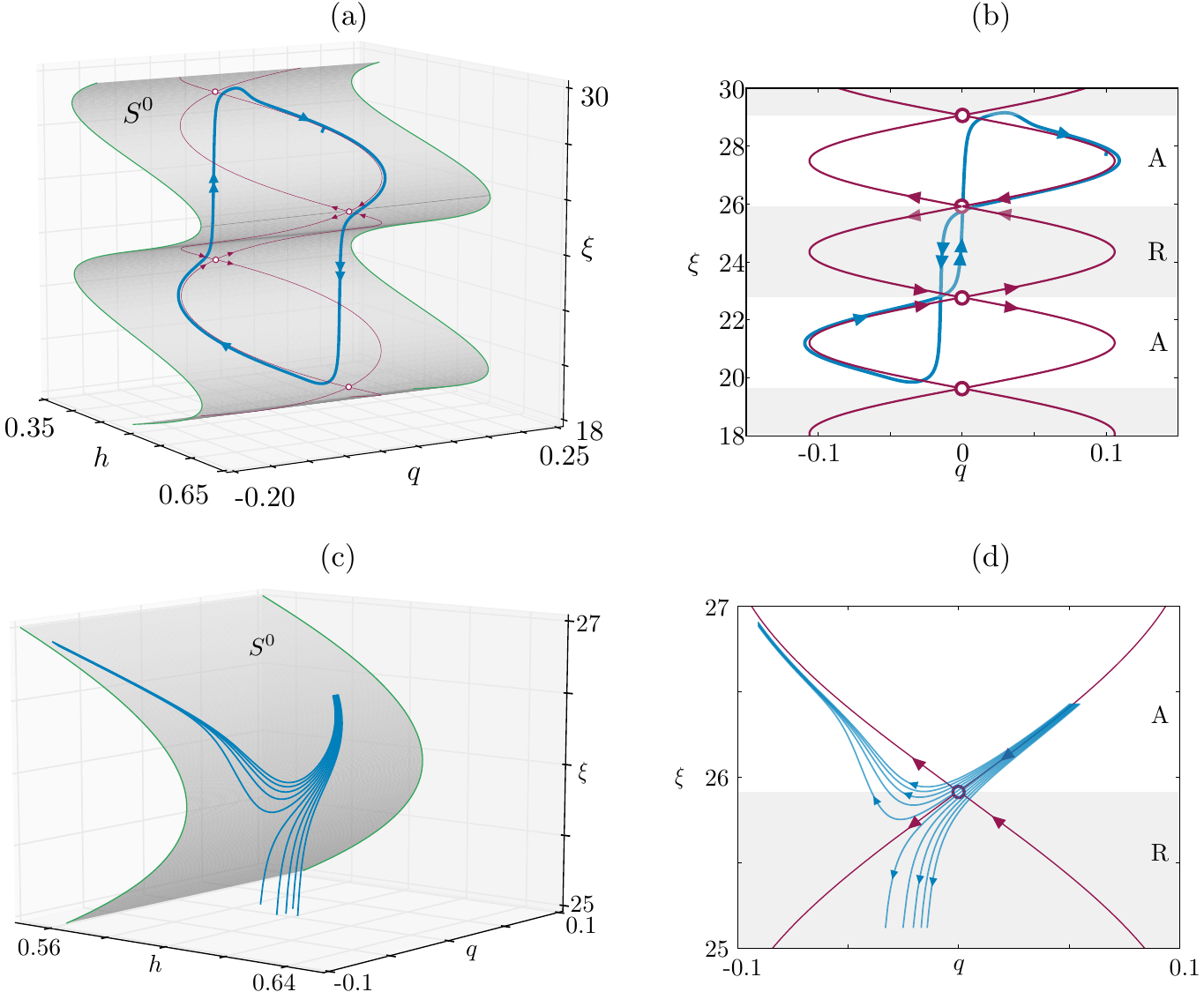}
  \caption{Additional features of the folded-saddle scenario. In (a)--(b) we repeat
  the simulation of Fig.~\ref{fig:foldedSingCanards} with $\epsi = 3.6 \cdot 10^{-3}$,
  finding a periodic solution displaying a \emph{jump-on canard}. (a): Projection of the
  trajectory of the full spatio-temporal model on the $(h,q,\xi)$ space (blue), where
  we also plot the critical manifold $S^0$ (grey) and singular canards on $S^0$
  (red). (b): Projection on the \edits{$(q,\xi)$} plane with attracting (A) and repelling (R)
  sheets of $S^0$. In (c)--(d) we show a family of orbit segments passing near the
  folded saddle and displaying sensitivity to initial conditions, separating
  orbits jumping up towards (A) or jumping down towards another attracting sheet of
  $S^0$ (also marked as (A) in panel (b)).}
  \label{fig:foldedSaddleSuppl}
\end{figure*}

We have confirmed these predictions using the full model~\eqref{eq:extendedNeuralField} 
with the heterogeneous synaptic kernel 
\begin{equation}\label{eq:heterokernel}
 W(x,y) = \frac{1}{2} e^{-|x-y|} \Big(a + b \cos \frac{y}{\Lambda} \Big),
\end{equation}
where $a,b \geq 0$, $\Lambda > 0$
and the firing rate function
\begin{equation}
  f(u) = (1 + e^{-\mu u})^{-1}.\label{eq:firingRateSupp}
\end{equation}
For $\mu \gg 1$ this sigmoidal function approximates a Heaviside firing rate employed
in the theory. We use the spectral algorithm developed in Ref.~\cite{Rankin:2014bz}
to solve the resulting equations. System~\eqref{eq:neuralFieldModel}, where $h$ is a
fixed parameter, admits branches of localized steady states arranged in a
characteristic \emph{snakes-and-ladders} bifurcation structure exhibiting countably
many folds at which $\psi'(\xi_*) = 0$ (Fig.~\ref{fig:figs1-2}(d)). 

\subsection{Spatio-temporal folded-saddle canards}

We first consider the uncoupled case with $\alpha=0.5$, $\beta=\gamma=0$, which leads
to spatio-temporal folded-saddle canards.
Figure~\ref{fig:foldedSingCanards}(a) shows the solution of the full spatial
system~\eqref{eq:extendedNeuralField} in the form of a space-time plot while
Fig.~\ref{fig:foldedSingCanards}(b) shows the same results but projected onto the
\edits{$(h,q,\xi)$} space (blue curve), compared with the singular canards
of~\eqref{eq:slowFlow} (red curves). For reference we
plot $S^0$ in grey. In Fig.~\ref{fig:foldedSingCanards}(c) we show a projection onto
the $(\xi,q)$ plane, where we indicate folded saddles (open circles) and the
attracting (A) and repelling (R) sheets of $S^0$. For these parameters the theory
predicts the presence of a folded-saddle spatio-temporal canard in
system~\eqref{eq:extendedNeuralField} and the projection indeed displays behavior
typical of folded-saddle singularities in ODEs: the orbit follows the upper
attracting sheet, passes the folded singularity from right to left
(Fig.~\ref{fig:foldedSingCanards}(c)) and then continues near a
repelling sheet of $S^0$ for an $O(1)$ time, before a fast (anterior) jump leads to
the lower attracting sheet; the orbit returns to the upper attracting sheet with a
second (posterior) fast jump. 

Since the latter jump occurs near a folded saddle, this opens the possibility of a
\emph{jump-on canard} segment, in which the orbit jumps and lands on the upper
repelling sheet of $S^0$ before returning to the upper attracting one. We observe
this behavior in canard cycles obtained with slightly different parameter values, as
reported in Figs.~\ref{fig:foldedSaddleSuppl}(a)--(b): the solution is
periodic and passes near four different folded-saddle singularities, marked with red
circles in Fig.~\ref{fig:foldedSaddleSuppl}(b); this trajectory contains one clear
canard segment near the topmost folded saddle; this segment is a jump-on canard, as
the orbit makes a fast upward jump and then follows directly a repelling segment
along the maximal canard of the folded saddle. The trajectory then passes near the
other folded saddles, without displaying a clear canard segment.

In Figs.~\ref{fig:foldedSaddleSuppl}(c)--(d) we present a family of solutions of
Eqs.~\eqref{eq:extendedNeuralField} for different initial conditions near an
attracting sheet of $S^0$. This experiment explains the sensitivity documented in
Figs.~\ref{fig:figs1-2}(a)--(c), and highlights the transition through the canard in the
folded-saddle case. This corresponds---modulo a change of direction near the
repelling sheets of $S^0$---to a perturbation of the stable manifold of the folded
saddle (as a saddle equilibrium of the DRS). Indeed,
trajectories approach the canard and follow it past the folded saddle; the
trajectories are then repelled and jump to a lower or upper attracting sheet of
$S^0$, depending on their initial condition. In Fig.~\ref{fig:foldedSaddleSuppl}(d)
we plot the singular canards associated with
this folded saddle: the true canard (from A to R) and the so-called ``false'' (or
\textit{faux}) canard (from R to A), both shown in red. In this scenario, the true
canard plays the role of a separatrix between trajectories that jump upwards,
following the faux canard, and downwards, towards a different attracting sheet of
$S^0$.

\subsection{Spatio-temporal folded-node canards}

\begin{figure*}
  \includegraphics[width=\textwidth]{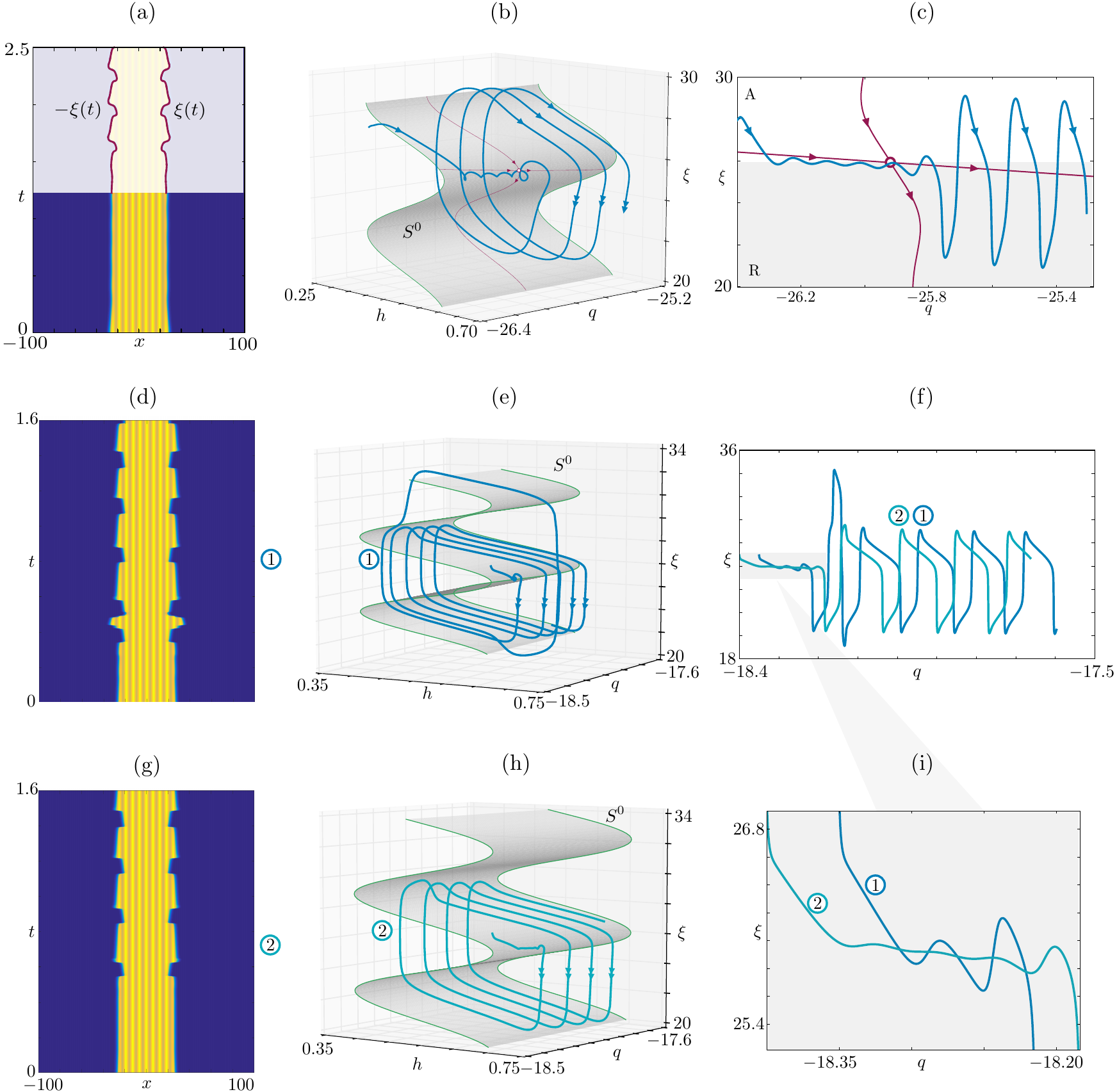}
  \caption{Examples of solutions containing spatio-temporal canards of
  folded-node type. Parameters are as in Fig.~\ref{fig:foldedSingCanards}, except
  $\alpha=\gamma=1$, $\beta=0$, for which the theory predicts folded-node canards. In
  (a)--(c) we plot the full spatio-temporal solution (a), its projection on the
  $(h,q,\xi)$ space (b) and on the \edits{$(q,\xi)$} plane (c); the latter show oscillations
  typical of folded-node canards, and therefore correspond to spatio-temporal
  folded-node canards in the neural field model. In (d)--(i) we set $\epsi = 3.6
  \cdot 10^{-3}$, $\alpha = 1$, $\gamma=0.7$ and retain all other parameter values;
  when initial conditions are varied slightly, a variable number of small
  oscillations is found near the folded node, as expected from the ODE theory.
  We set $q(0) = -18.35$ (label 1 in (d), (e), (f), (i)) and $q(0) = -18.40$
  (label 2 in (f), (g), (h), (i)). (f): Projections on the \edits{$(q,\xi)$} plane, revealing an
  initial drift near the folded node, during which trajectory 1 (2) displays 3 (5)
  small-amplitude oscillations around the folded singularity (see inset (i)).}
  \label{fig:multiOscillationsFoldedNode}
\end{figure*}

We next repeat our numerical analysis for the fully coupled
system~\eqref{eq:extendedNeuralField} when $\alpha=1$, $\beta=0$, $\gamma=1$ for
which the theory predicts spatio-temporal canards of folded-node type
(Fig.~\ref{fig:multiOscillationsFoldedNode}(a)--(c)). The
folded-node scenario is richer than the folded-saddle one: first, solutions
containing canard segments exist for $O(1)$ ranges of initial conditions
and parameter values; second, there are many more possible waveforms due
to the existence of a funnel region around the folded-node singularity that
induces a rotation of the trajectories as they pass through it; this effect
is clearly visible in Fig.~\ref{fig:multiOscillationsFoldedNode}(b)--(c) as
small amplitude spiraling motion in the vicinity of the fold, and rather less
clearly as the minute oscillations for $t \in [0, 1.5]$ in
Fig.~\ref{fig:multiOscillationsFoldedNode}(a). As initial conditions change, the
number of these small (subthreshold) oscillations in the funnel region
varies and this phenomenon defines rotation sectors near $S^0$. The boundaries
between different rotation sectors correspond to canard solutions generating
mixed-mode dynamics in the system. For fixed parameter values, the maximum
number of subthreshold oscillations is given by the eigenvalue ratio of the
folded node \cite{Desroches2012}, seen as an equilibrium of
Eqs.~\eqref{eq:desingSlowFlow}. Thus trajectories with different initial
conditions will be trapped in the funnel and pass near the folded node while
making different numbers of subthreshold oscillations, thereby encoding the possible
waveforms in this regime.

We exemplify this behavior in Figs.~\ref{fig:multiOscillationsFoldedNode}(d)--(i) by
time-stepping~\eqref{eq:extendedNeuralField} with slightly different initial conditions, 
close to a folded node, when $\alpha=1$, $\beta=0$, $\gamma=0.7$. In the experiment
under consideration we pre-computed a stationary pattern $u_0(x)$ for the neural
field equation with constant firing rate threshold, $h = 0.57$, $\dot h=0$, that is,
we select a stationary state on $S^0$. We then perturb this state and compute two
trajectories, with initial conditions close to the folded node by setting $u(x,0) =
u_0(x)$, $h(0) = 0.58$, $q(0) = -18.40$ (label 1) and $q(0) = -18.35$ (label 2).
\edits{Figures~\ref{fig:multiOscillationsFoldedNode}(d,g)} show the corresponding space-time
evolution $u(x,t)$ while Figs.~\ref{fig:multiOscillationsFoldedNode}(e,h) show the
corresponding trajectories
in $(h,q,\xi)$ space. Panel (c) and the enlargement in (f) show the projections of these
trajectories on the $(q,\xi)$ plane. The trajectories are initially close
and exhibit the drifting and spiralling motion predicted by the theory~\cite{Desroches2012},
with respectively three and five subthreshold oscillations near the folded node.
After an initial transient, in which trajectory 1 visits the upper attracting sheet
of $S^0$, both trajectories wrap clockwise around the middle and bottom attracting
sheets of $S^0$ (Figs.~\ref{fig:multiOscillationsFoldedNode}(e,h)). Both also
display jump-on canard segments at every turn, in the vicinity of the left boundary of the 
upper repelling sheet of $S^0$, although these become less pronounced as time increases.

\section{Neural fields posed on a sphere}

\begin{figure*}
  \includegraphics{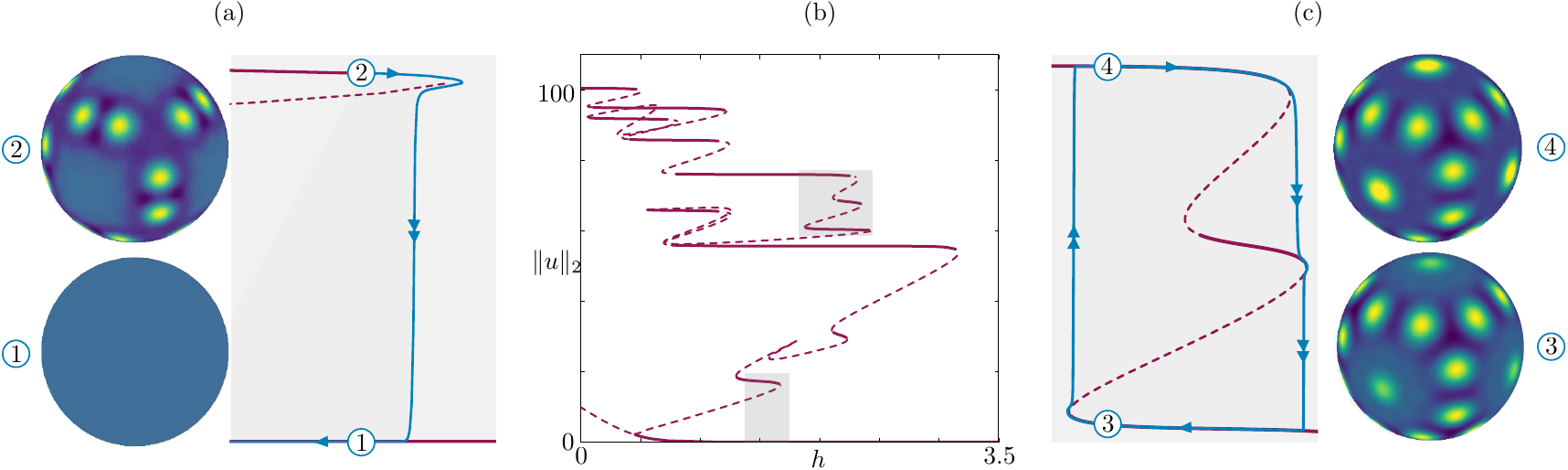}
  \caption{Spatio-temporal canards of folded-saddle type occurring in a neural field model
    posed on a spherical domain using $h$ as the continuation parameter. (b) Bifurcation diagram of steady states with
    octahedral symmetry in the system~\eqref{eq:sphereModelSupp}.
    (a,c) Bifurcation diagram (red), orbit (blue) and representative patterns
    obtained when $h$ varies slowly through (a) a low-lying fold, and (c) one of the
    higher folds.  When the evolution of $h$ is decoupled from $u$, we observe spatio-temporal
    canards of folded-saddle type (see the animations
    \href{run:anc/canardCycle.mp4}{canardCycle.mp4},
    \href{run:anc/canardDown.mp4}{canardDown.mp4}, 
    \href{run:anc/canardUp.mp4}{canardUp.mp4} for further details).
    }
  \label{fig:spheresBifDiag}
\end{figure*}

We have also studied neural field models posed on a more realistic spherical domain
and identified spatio-temporal canards with octahedral
symmetry where interfaces are no longer points but curves in 3D. The above theory does
not readily generalize to this setting but we nevertheless successfully tested its
predictions in the folded-saddle case, when $u$ and $h$ are decoupled. As shown in
Fig.~\ref{fig:spheresBifDiag}, the model displays orbits with canard segments (and
canard cycles). In this case the system with constant $h$ admits an intricate
bifurcation diagram (not shown), where coexisting stable states with octahedral,
icosahedral and rotational symmetry are interconnected via symmetry-breaking
bifurcations and saddle-node bifurcations. 

The calculations for neural fields posed on the unit sphere
were performed using a neural field model with a constant threshold crossing $h$,
\begin{align} \label{eq:sphereModelSupp}
  \partial_t u(\vect{x},t) &= -u(\vect{x},t)\\
\nonumber  &+ \kappa \int_{\sph} W\big( \langle \vect{x},\vect{y} \rangle \big) f\big(
  u(\vect{y},t) - h\big) \, d\sigma(\vect{y}), 
\end{align}
where $\vect{x} \in \sph = \{ \vect{z} \in \RSet^3 \colon |
\vect{z} | = 1 \}$ and the integral is over $\sph$.
In this integro-differential equation the kernel $W$ models the synaptic wiring
between two points $\vect{x}, \vect{y}$ on the surface of a sphere;
we assume that this wiring depends solely on the great-circle distance (geodesic)
between $\vect{x}$ and $\vect{y}$, hence the dependence on the scalar product $\langle
\vect{x},\vect{y} \rangle$. We use an excitatory-inhibitory Gaussian synaptic kernel
\begin{equation}\label{eq:sphereKernel}
W(\xi) = A_1 \exp(-\xi^2 /B_1) - A_2 \exp(-\xi^2 /B_2).
\end{equation}
Stationary patterned states of \eqref{eq:sphereModelSupp} were continued in the parameter
$h$ using a Nystr\"om scheme, combined with standard path-following techniques \edits{as well as} 
high-order, highly efficient, icosahedral- or tetrahedral-invariant quadrature schemes.
A comprehensive study of branches of patterned states supported by this model, their
symmetries and stability, as well as the properties of the numerical scheme will be
described in a separate publication~\cite{Avitabile:2016}. A sample result showing
a branch of states with octahedral symmetry
 is reported in Fig.~\ref{fig:spheresBifDiag}(b). Solutions with
this symmetry bifurcate transcritically from the homogeneous steady state, and then
undergo a sequence of saddle-nodes and symmetry-breaking bifurcations shown in the figure.

We are interested in testing the predictions of the theory developed for 1D domains for
more realistic cortical surfaces. For physical domains in higher dimensions, it is
possible to reduce the equations as for 1D domains, but the reduction
is still a spatially-extended dynamical system. In 1D, the activity set
is given by $\mathcal{A}(t) = [-\xi(t), \xi(t)] \in \RSet$ and, differentiating one
of the threshold conditions, say $u(\xi(t),t) = h(t)$, we obtain
\begin{equation}\label{eq:evolEq1DSupp}
  \partial_x u(\xi(t),t) \dot \xi(t) = h(t) + \dot h(t) - \int_{-\xi(t)}^{\xi(t)}
   W(x,y)\, dy
\end{equation}
which is an evolution equation for the scalar variable $\xi$. To extend this procedure
to the sphere, we assume that the activity set 
$
\mathcal{A}(t) = \{ \vect{x} \in \sph \colon u(\vect{x},t) \geq h(t) \}
$
has a boundary which can be parameterized as follows,
\begin{align*}
  \partial \mathcal{A}(t) & = \bigcup_{k = 1}^K \mathcal{C}_k(t),\\
  \mathcal{C}_k(t) & = \{ \vect{x} \in \sph \colon \vx = \vxi_k(s,t), \; s \in [0,2\pi) \},
\end{align*}
where the functions $\{ \vxi_k \}$ are $2\pi$-periodic and smooth in the variable $s$. In other
words, we assume that the boundary of the activity set on the spherical domain is the
union of $K$ disjoint curves on the spherical surface $\sph$. We seek
evolution equations for the functions $\{ \vxi_k \}$. Since the solution $u(\vx,t)$ crosses
the threshold $h(t)$ on each of the curves $\mathcal{C}_k$, we differentiate the threshold
condition $u(\vxi_k(s,t),t) = h(t)$ with respect to $t$ to obtain
\begin{align}
\hspace*{-1cm}\label{eq:evolEqSupp}\big\langle \grad u \big(\vxi_k(s,t),t\big),& \; \partial_t \vxi_k(s,t) \big\rangle = h(t) + \dot h(t)\\
\nonumber& - \int_{\mathcal{A}(t)} W\big( \langle \vxi_k(s,t), \vy \rangle \big) \, d\sigma(\vy),\\ 
\nonumber& s \in [0,2\pi),\quad k = 1,\ldots,K,\\
\vxi_k(0,t) & = \vxi_k(2\pi,t), \quad k = 1,\ldots,K,
\label{eq:BCsSupp}
\end{align}
where the gradient is in spherical coordinates. It can be shown that, under suitable assumptions
on the kernel, the inner product on the left hand side and the surface integral on the right hand
side of~\eqref{eq:evolEqSupp} can be written~\cite{Coombes2012aa,Coombes2013aa} in terms of line
integrals over the closed curves $\mathcal{C}_k(t)$.
The system~\eqref{eq:evolEqSupp}--\eqref{eq:BCsSupp} is therefore closed and represents a generalization
of~\eqref{eq:evolEq1DSupp}. In this case, however, the state variables are the functions
$\{ \vxi_k(s) \}$, as opposed to the scalar $\xi$, and a canard theory for this system
is currently unavailable.

We can, however, simulate the system~\eqref{eq:sphereModelSupp} or the
system~\eqref{eq:evolEqSupp}--\eqref{eq:BCsSupp} numerically and search for evidence of spatio-temporal
canards. More precisely, we have performed numerical experiments to test the
robustness of the 1D theory to
\begin{enumerate}
  \item Changes in the geometry of the problem: the spherical model includes
    curvature effects via the great-circle distance $\langle \vect{x}, \vect{y} \rangle$
    between points $\vect{x}$, $\vect{y}$ on the spherical cortex (see
    Eq.~\eqref{eq:sphereModelSupp}).
  \item Changes in the synaptic connectivity function: the
    kernel~\eqref{eq:sphereKernel} is different from that used in the 1D
    computations; in particular, the kernel~\eqref{eq:sphereKernel} is
    excitatory-inhibitory and homogeneous while kernel~\eqref{eq:heterokernel} is
    purely excitatory and heterogeneous.
  \item Changes in the firing rate function: the theory is valid for a Heaviside
    firing rate which is approximated in the 1D simulations by a steep
    sigmoid (Eq.~\eqref{eq:firingRateSupp} with $\mu=50$); in the spherical simulations
    we employ a shallow firing rate ($\mu = 8$).
  \item Changes in the evolution equation of the firing threshold $h$: in the
    spherical simulations, $h$ evolves slowly and independently from $u$, but not
    harmonically:
    \begin{equation}\label{eq:hEvolution}
      h(t) = 
      \begin{cases}
	\epsi t + h_0           & 0 \leq t    < (h_1 - h_0)/\epsi \\
	-\epsi t + 2h_1 - h_0   & t \geq (h_1 - h_0)/\epsi,
      \end{cases}
    \end{equation}
    where $h_1$ is a fold point in the bifurcation diagram (located using standard
    bifurcation analysis techniques) and $h_0<h_1$. Consequently, $h$ undergoes a
    slow linear increase up to the fold, followed by a slow linear decrease.
\end{enumerate}
In each case we found that the qualitative predictions of the 1D theory carried over to this
much more complicated situation.
\section{Conclusions and perspectives}
To the best of our knowledge, this article presents the first
theory for folded-singularity temporal canards in a spatially-extended system. This
result paves the way towards a systematic study of spatio-temporal mixed-mode
oscillations (MMOs) in spatially-extended systems, 
\edits{with the view of explaining the origin of MMOs observed in spatio-temporal
signals modelling spike-frequency adaptation and synaptic
depression~\cite{Folias2005aaa}.}
\edits{The spatio-temporal structures discussed here are also directly relevant to neural mass and
connectomic models, in which a discrete connectomic matrix replaces the heterogeneous
kernel $W$~\cite{Haimovici2013}: canard structures in
these models would offer a rigorous explanation of the brutal transitions observed,
for instance, in models of partial epilepsy~\cite{proix2014}}. There is a general
consensus that spike (and more generally burst) timings, durations and rates are
involved in information coding in the brain~\cite{borst1999}.  Being able to identify
boundaries (represented by spatio-temporal canards) between different activity
regimes (e.g.\ spiking/bursting or mixed-mode oscillations with different signatures)
may shed further light on the transmission of information in
the brain.
\newline
\textbf{Acknowledgement}: 
This work was supported in part by the Engineering and Physical Sciences Research
Council under grant EP/P510993/1 (DA) and by the National Science
Foundation under grant DMS-1613132 (EK). DA thanks Luke Wood and Oliver Smith
\edits{for their work on neural field models during their final-year undergraduate
dissertations.}

\noindent\textbf{Author contributions}: DA and MD contributed equally to this work.


\bibliographystyle{apsrev4-1} 
\bibliography{referencesFinal}

\begin{thebibliography}{44}%
\makeatletter
\providecommand \@ifxundefined [1]{%
 \@ifx{#1\undefined}
}%
\providecommand \@ifnum [1]{%
 \ifnum #1\expandafter \@firstoftwo
 \else \expandafter \@secondoftwo
 \fi
}%
\providecommand \@ifx [1]{%
 \ifx #1\expandafter \@firstoftwo
 \else \expandafter \@secondoftwo
 \fi
}%
\providecommand \natexlab [1]{#1}%
\providecommand \enquote  [1]{``#1''}%
\providecommand \bibnamefont  [1]{#1}%
\providecommand \bibfnamefont [1]{#1}%
\providecommand \citenamefont [1]{#1}%
\providecommand \href@noop [0]{\@secondoftwo}%
\providecommand \href [0]{\begingroup \@sanitize@url \@href}%
\providecommand \@href[1]{\@@startlink{#1}\@@href}%
\providecommand \@@href[1]{\endgroup#1\@@endlink}%
\providecommand \@sanitize@url [0]{\catcode `\\12\catcode `\$12\catcode
  `\&12\catcode `\#12\catcode `\^12\catcode `\_12\catcode `\%12\relax}%
\providecommand \@@startlink[1]{}%
\providecommand \@@endlink[0]{}%
\providecommand \url  [0]{\begingroup\@sanitize@url \@url }%
\providecommand \@url [1]{\endgroup\@href {#1}{\urlprefix }}%
\providecommand \urlprefix  [0]{URL }%
\providecommand \Eprint [0]{\href }%
\providecommand \doibase [0]{http://dx.doi.org/}%
\providecommand \selectlanguage [0]{\@gobble}%
\providecommand \bibinfo  [0]{\@secondoftwo}%
\providecommand \bibfield  [0]{\@secondoftwo}%
\providecommand \translation [1]{[#1]}%
\providecommand \BibitemOpen [0]{}%
\providecommand \bibitemStop [0]{}%
\providecommand \bibitemNoStop [0]{.\EOS\space}%
\providecommand \EOS [0]{\spacefactor3000\relax}%
\providecommand \BibitemShut  [1]{\csname bibitem#1\endcsname}%
\let\auto@bib@innerbib\@empty
\bibitem [{\citenamefont {Bressloff}(2012)}]{Bressloff2012o}%
  \BibitemOpen
  \bibfield  {author} {\bibinfo {author} {\bibfnamefont {P.~C.}\ \bibnamefont
  {Bressloff}},\ }\href@noop {} {\bibfield  {journal} {\bibinfo  {journal} {J.
  Phys. A}\ }\textbf {\bibinfo {volume} {45}},\ \bibinfo {pages} {{033001}}
  (\bibinfo {year} {2012})}\BibitemShut {NoStop}%
\bibitem [{\citenamefont {Bressloff}(2014)}]{Bressloff:2014cm}%
  \BibitemOpen
  \bibfield  {author} {\bibinfo {author} {\bibfnamefont {P.~C.}\ \bibnamefont
  {Bressloff}},\ }\href@noop {} {\emph {\bibinfo {title} {Waves in Neural
  Media}}}\ (\bibinfo  {publisher} {Springer},\ \bibinfo {address} {New York,
  NY},\ \bibinfo {year} {2014})\BibitemShut {NoStop}%
\bibitem [{\citenamefont {Ermentrout}\ and\ \citenamefont
  {Terman}(2010)}]{Ermentrout:2010cga}%
  \BibitemOpen
  \bibfield  {author} {\bibinfo {author} {\bibfnamefont {G.~B.}\ \bibnamefont
  {Ermentrout}}\ and\ \bibinfo {author} {\bibfnamefont {D.~H.}\ \bibnamefont
  {Terman}},\ }\href@noop {} {\emph {\bibinfo {title} {{Mathematical
  Foundations of Neuroscience}}}}\ (\bibinfo  {publisher} {Springer},\ \bibinfo
  {address} {New York},\ \bibinfo {year} {2010})\BibitemShut {NoStop}%
\bibitem [{\citenamefont {Wilson}\ and\ \citenamefont
  {Cowan}(1972)}]{Wilson1972aa}%
  \BibitemOpen
  \bibfield  {author} {\bibinfo {author} {\bibfnamefont {H.~R.}\ \bibnamefont
  {Wilson}}\ and\ \bibinfo {author} {\bibfnamefont {J.~D.}\ \bibnamefont
  {Cowan}},\ }\href@noop {} {\bibfield  {journal} {\bibinfo  {journal}
  {Biophys. J.}\ }\textbf {\bibinfo {volume} {12}},\ \bibinfo {pages} {1}
  (\bibinfo {year} {1972})}\BibitemShut {NoStop}%
\bibitem [{\citenamefont {Amari}(1975)}]{Amari1975aa}%
  \BibitemOpen
  \bibfield  {author} {\bibinfo {author} {\bibfnamefont {S.-I.}\ \bibnamefont
  {Amari}},\ }\href@noop {} {\bibfield  {journal} {\bibinfo  {journal} {Biol.
  Cybern.}\ }\textbf {\bibinfo {volume} {17}},\ \bibinfo {pages} {211}
  (\bibinfo {year} {1975})}\BibitemShut {NoStop}%
\bibitem [{\citenamefont {Folias}\ and\ \citenamefont
  {Bressloff}(2005{\natexlab{a}})}]{Folias2005aa}%
  \BibitemOpen
  \bibfield  {author} {\bibinfo {author} {\bibfnamefont {S.}~\bibnamefont
  {Folias}}\ and\ \bibinfo {author} {\bibfnamefont {P.}~\bibnamefont
  {Bressloff}},\ }\href@noop {} {\bibfield  {journal} {\bibinfo  {journal}
  {Phys. Rev. Lett.}\ }\textbf {\bibinfo {volume} {95}},\ \bibinfo {pages}
  {208107} (\bibinfo {year} {2005}{\natexlab{a}})}\BibitemShut {NoStop}%
\bibitem [{\citenamefont {Richardson}\ \emph {et~al.}(2005)\citenamefont
  {Richardson}, \citenamefont {Schiff},\ and\ \citenamefont
  {Gluckman}}]{Richardson:2005cs}%
  \BibitemOpen
  \bibfield  {author} {\bibinfo {author} {\bibfnamefont {K.~A.}\ \bibnamefont
  {Richardson}}, \bibinfo {author} {\bibfnamefont {S.~J.}\ \bibnamefont
  {Schiff}}, \ and\ \bibinfo {author} {\bibfnamefont {B.~J.}\ \bibnamefont
  {Gluckman}},\ }\href@noop {} {\bibfield  {journal} {\bibinfo  {journal}
  {Phys. Rev. Lett.}\ }\textbf {\bibinfo {volume} {94}},\ \bibinfo {pages}
  {028103} (\bibinfo {year} {2005})}\BibitemShut {NoStop}%
\bibitem [{\citenamefont {Huang}\ \emph {et~al.}(2004)\citenamefont {Huang},
  \citenamefont {Troy}, \citenamefont {Yang}, \citenamefont {Ma}, \citenamefont
  {Laing}, \citenamefont {Schiff},\ and\ \citenamefont {Wu}}]{Huang:2004kw}%
  \BibitemOpen
  \bibfield  {author} {\bibinfo {author} {\bibfnamefont {X.}~\bibnamefont
  {Huang}}, \bibinfo {author} {\bibfnamefont {W.~C.}\ \bibnamefont {Troy}},
  \bibinfo {author} {\bibfnamefont {Q.}~\bibnamefont {Yang}}, \bibinfo {author}
  {\bibfnamefont {H.}~\bibnamefont {Ma}}, \bibinfo {author} {\bibfnamefont
  {C.~R.}\ \bibnamefont {Laing}}, \bibinfo {author} {\bibfnamefont {S.~J.}\
  \bibnamefont {Schiff}}, \ and\ \bibinfo {author} {\bibfnamefont {J.-Y.}\
  \bibnamefont {Wu}},\ }\href@noop {} {\bibfield  {journal} {\bibinfo
  {journal} {J. Neurosci.}\ }\textbf {\bibinfo {volume} {24}},\ \bibinfo
  {pages} {9897} (\bibinfo {year} {2004})}\BibitemShut {NoStop}%
\bibitem [{\citenamefont {Gonz{\'a}lez-Ram{\'\i}rez}\ \emph
  {et~al.}(2015)\citenamefont {Gonz{\'a}lez-Ram{\'\i}rez}, \citenamefont
  {Ahmed}, \citenamefont {Cash}, \citenamefont {Wayne},\ and\ \citenamefont
  {Kramer}}]{GonzalezRamirez:2015gk}%
  \BibitemOpen
  \bibfield  {author} {\bibinfo {author} {\bibfnamefont {L.~R.}\ \bibnamefont
  {Gonz{\'a}lez-Ram{\'\i}rez}}, \bibinfo {author} {\bibfnamefont {O.~J.}\
  \bibnamefont {Ahmed}}, \bibinfo {author} {\bibfnamefont {S.~S.}\ \bibnamefont
  {Cash}}, \bibinfo {author} {\bibfnamefont {C.~E.}\ \bibnamefont {Wayne}}, \
  and\ \bibinfo {author} {\bibfnamefont {M.~A.}\ \bibnamefont {Kramer}},\
  }\href@noop {} {\bibfield  {journal} {\bibinfo  {journal} {PLoS Comput.
  Biol.}\ }\textbf {\bibinfo {volume} {11}},\ \bibinfo {pages} {e1004065}
  (\bibinfo {year} {2015})}\BibitemShut {NoStop}%
\bibitem [{\citenamefont {Steyn-Ross}\ \emph {et~al.}(2003)\citenamefont
  {Steyn-Ross}, \citenamefont {Steyn-Ross}, \citenamefont {Sleigh},\ and\
  \citenamefont {Whiting}}]{SteynRoss:2003ep}%
  \BibitemOpen
  \bibfield  {author} {\bibinfo {author} {\bibfnamefont {M.~L.}\ \bibnamefont
  {Steyn-Ross}}, \bibinfo {author} {\bibfnamefont {D.~A.}\ \bibnamefont
  {Steyn-Ross}}, \bibinfo {author} {\bibfnamefont {J.~W.}\ \bibnamefont
  {Sleigh}}, \ and\ \bibinfo {author} {\bibfnamefont {D.~R.}\ \bibnamefont
  {Whiting}},\ }\href@noop {} {\bibfield  {journal} {\bibinfo  {journal} {Phys.
  Rev. E}\ }\textbf {\bibinfo {volume} {68}},\ \bibinfo {pages} {021902}
  (\bibinfo {year} {2003})}\BibitemShut {NoStop}%
\bibitem [{\citenamefont {Camperi}\ and\ \citenamefont
  {Wang}(1998)}]{Camperi:1998ji}%
  \BibitemOpen
  \bibfield  {author} {\bibinfo {author} {\bibfnamefont {M.}~\bibnamefont
  {Camperi}}\ and\ \bibinfo {author} {\bibfnamefont {X.-J.}\ \bibnamefont
  {Wang}},\ }\href@noop {} {\bibfield  {journal} {\bibinfo  {journal} {J.
  Comput. Neurosci.}\ }\textbf {\bibinfo {volume} {5}},\ \bibinfo {pages} {383}
  (\bibinfo {year} {1998})}\BibitemShut {NoStop}%
\bibitem [{\citenamefont {Beno\^{i}t}\ \emph {et~al.}(1981)\citenamefont
  {Beno\^{i}t}, \citenamefont {Callot}, \citenamefont {Diener},\ and\
  \citenamefont {Diener}}]{Benoit1981}%
  \BibitemOpen
  \bibfield  {author} {\bibinfo {author} {\bibfnamefont {E.}~\bibnamefont
  {Beno\^{i}t}}, \bibinfo {author} {\bibfnamefont {J.-L.}\ \bibnamefont
  {Callot}}, \bibinfo {author} {\bibfnamefont {F.}~\bibnamefont {Diener}}, \
  and\ \bibinfo {author} {\bibfnamefont {M.}~\bibnamefont {Diener}},\
  }\href@noop {} {\bibfield  {journal} {\bibinfo  {journal} {Collect. Math.}\
  }\textbf {\bibinfo {volume} {32}},\ \bibinfo {pages} {37} (\bibinfo {year}
  {1981})}\BibitemShut {NoStop}%
\bibitem [{\citenamefont {Krupa}\ and\ \citenamefont
  {Szmolyan}(2001)}]{Krupa2001}%
  \BibitemOpen
  \bibfield  {author} {\bibinfo {author} {\bibfnamefont {M.}~\bibnamefont
  {Krupa}}\ and\ \bibinfo {author} {\bibfnamefont {P.}~\bibnamefont
  {Szmolyan}},\ }\href@noop {} {\bibfield  {journal} {\bibinfo  {journal} {J.
  Differ. Equations}\ }\textbf {\bibinfo {volume} {174}},\ \bibinfo {pages}
  {312} (\bibinfo {year} {2001})}\BibitemShut {NoStop}%
\bibitem [{\citenamefont {Desroches}\ \emph
  {et~al.}(2013{\natexlab{a}})\citenamefont {Desroches}, \citenamefont
  {Krupa},\ and\ \citenamefont {Rodrigues}}]{Desroches2013b}%
  \BibitemOpen
  \bibfield  {author} {\bibinfo {author} {\bibfnamefont {M.}~\bibnamefont
  {Desroches}}, \bibinfo {author} {\bibfnamefont {M.}~\bibnamefont {Krupa}}, \
  and\ \bibinfo {author} {\bibfnamefont {S.}~\bibnamefont {Rodrigues}},\
  }\href@noop {} {\bibfield  {journal} {\bibinfo  {journal} {J. Math. Biol.}\
  }\textbf {\bibinfo {volume} {67}},\ \bibinfo {pages} {989} (\bibinfo {year}
  {2013}{\natexlab{a}})}\BibitemShut {NoStop}%
\bibitem [{\citenamefont {Mitry}\ \emph {et~al.}(2013)\citenamefont {Mitry},
  \citenamefont {McCarthy}, \citenamefont {Kopell},\ and\ \citenamefont
  {Wechselberger}}]{Mitry2013}%
  \BibitemOpen
  \bibfield  {author} {\bibinfo {author} {\bibfnamefont {J.}~\bibnamefont
  {Mitry}}, \bibinfo {author} {\bibfnamefont {M.}~\bibnamefont {McCarthy}},
  \bibinfo {author} {\bibfnamefont {N.}~\bibnamefont {Kopell}}, \ and\ \bibinfo
  {author} {\bibfnamefont {M.}~\bibnamefont {Wechselberger}},\ }\href@noop {}
  {\bibfield  {journal} {\bibinfo  {journal} {J. Math. Neurosci.}\ }\textbf
  {\bibinfo {volume} {3}},\ \bibinfo {pages} {1} (\bibinfo {year}
  {2013})}\BibitemShut {NoStop}%
\bibitem [{\citenamefont {Moehlis}(2006)}]{Moehlis2006}%
  \BibitemOpen
  \bibfield  {author} {\bibinfo {author} {\bibfnamefont {J.}~\bibnamefont
  {Moehlis}},\ }\href@noop {} {\bibfield  {journal} {\bibinfo  {journal} {J.
  Math. Biol.}\ }\textbf {\bibinfo {volume} {52}},\ \bibinfo {pages} {141}
  (\bibinfo {year} {2006})}\BibitemShut {NoStop}%
\bibitem [{\citenamefont {Kramer}\ \emph {et~al.}(2008)\citenamefont {Kramer},
  \citenamefont {Traub},\ and\ \citenamefont {Kopell}}]{Kramer2008}%
  \BibitemOpen
  \bibfield  {author} {\bibinfo {author} {\bibfnamefont {M.~A.}\ \bibnamefont
  {Kramer}}, \bibinfo {author} {\bibfnamefont {R.~D.}\ \bibnamefont {Traub}}, \
  and\ \bibinfo {author} {\bibfnamefont {N.~J.}\ \bibnamefont {Kopell}},\
  }\href@noop {} {\bibfield  {journal} {\bibinfo  {journal} {Phys. Rev. Lett.}\
  }\textbf {\bibinfo {volume} {101}},\ \bibinfo {pages} {68103} (\bibinfo
  {year} {2008})}\BibitemShut {NoStop}%
\bibitem [{\citenamefont {Rinzel}(1987)}]{Rinzel1986}%
  \BibitemOpen
  \bibfield  {author} {\bibinfo {author} {\bibfnamefont {J.}~\bibnamefont
  {Rinzel}},\ }in\ \href@noop {} {\emph {\bibinfo {booktitle} {Proc. Intern.
  Congr. Math.}}},\ Vol.\ \bibinfo {volume} {1-2}\ (\bibinfo  {publisher}
  {Amer. Math. Soc.},\ \bibinfo {address} {Providence, RI},\ \bibinfo {year}
  {1987})\ pp.\ \bibinfo {pages} {1578--1593}\BibitemShut {NoStop}%
\bibitem [{\citenamefont {Desroches}\ \emph {et~al.}(2012)\citenamefont
  {Desroches}, \citenamefont {Guckenheimer}, \citenamefont {Krauskopf},
  \citenamefont {Kuehn}, \citenamefont {Osinga},\ and\ \citenamefont
  {Wechselberger}}]{Desroches2012}%
  \BibitemOpen
  \bibfield  {author} {\bibinfo {author} {\bibfnamefont {M.}~\bibnamefont
  {Desroches}}, \bibinfo {author} {\bibfnamefont {J.}~\bibnamefont
  {Guckenheimer}}, \bibinfo {author} {\bibfnamefont {B.}~\bibnamefont
  {Krauskopf}}, \bibinfo {author} {\bibfnamefont {C.}~\bibnamefont {Kuehn}},
  \bibinfo {author} {\bibfnamefont {H.~M.}\ \bibnamefont {Osinga}}, \ and\
  \bibinfo {author} {\bibfnamefont {M.}~\bibnamefont {Wechselberger}},\
  }\href@noop {} {\bibfield  {journal} {\bibinfo  {journal} {SIAM Rev.}\
  }\textbf {\bibinfo {volume} {54}},\ \bibinfo {pages} {211} (\bibinfo {year}
  {2012})}\BibitemShut {NoStop}%
\bibitem [{\citenamefont {Desroches}\ \emph
  {et~al.}(2013{\natexlab{b}})\citenamefont {Desroches}, \citenamefont
  {Kaper},\ and\ \citenamefont {Krupa}}]{Desroches2013}%
  \BibitemOpen
  \bibfield  {author} {\bibinfo {author} {\bibfnamefont {M.}~\bibnamefont
  {Desroches}}, \bibinfo {author} {\bibfnamefont {T.~J.}\ \bibnamefont
  {Kaper}}, \ and\ \bibinfo {author} {\bibfnamefont {M.}~\bibnamefont
  {Krupa}},\ }\href@noop {} {\bibfield  {journal} {\bibinfo  {journal} {Chaos}\
  }\textbf {\bibinfo {volume} {23}},\ \bibinfo {pages} {046106} (\bibinfo
  {year} {2013}{\natexlab{b}})}\BibitemShut {NoStop}%
\bibitem [{\citenamefont {Gandhi}\ \emph {et~al.}(2016)\citenamefont {Gandhi},
  \citenamefont {Beaume},\ and\ \citenamefont {Knobloch}}]{Gandhi16}%
  \BibitemOpen
  \bibfield  {author} {\bibinfo {author} {\bibfnamefont {P.}~\bibnamefont
  {Gandhi}}, \bibinfo {author} {\bibfnamefont {C.}~\bibnamefont {Beaume}}, \
  and\ \bibinfo {author} {\bibfnamefont {E.}~\bibnamefont {Knobloch}},\ }in\
  \href@noop {} {\emph {\bibinfo {booktitle} {Nonlinear Dynamics: Materials,
  Theory and Experiments}}},\ \bibinfo {editor} {edited by\ \bibinfo {editor}
  {\bibfnamefont {M.}~\bibnamefont {Tlidi}}\ and\ \bibinfo {editor}
  {\bibfnamefont {M.~G.}\ \bibnamefont {Clerc}}}\ (\bibinfo  {publisher}
  {Springer},\ \bibinfo {address} {New York},\ \bibinfo {year} {2016})\ pp.\
  \bibinfo {pages} {303--316}\BibitemShut {NoStop}%
\bibitem [{\citenamefont {Coombes}(2005)}]{Coombes2005aa}%
  \BibitemOpen
  \bibfield  {author} {\bibinfo {author} {\bibfnamefont {S.}~\bibnamefont
  {Coombes}},\ }\href@noop {} {\bibfield  {journal} {\bibinfo  {journal} {Biol.
  Cyber.}\ }\textbf {\bibinfo {volume} {93}},\ \bibinfo {pages} {91} (\bibinfo
  {year} {2005})}\BibitemShut {NoStop}%
\bibitem [{\citenamefont {Laing}\ and\ \citenamefont
  {Troy}(2003)}]{Laing2003aa}%
  \BibitemOpen
  \bibfield  {author} {\bibinfo {author} {\bibfnamefont {C.~R.}\ \bibnamefont
  {Laing}}\ and\ \bibinfo {author} {\bibfnamefont {W.~C.}\ \bibnamefont
  {Troy}},\ }\href@noop {} {\bibfield  {journal} {\bibinfo  {journal} {SIAM J.
  Appl. Dyn. Sys.}\ }\textbf {\bibinfo {volume} {2}},\ \bibinfo {pages} {487}
  (\bibinfo {year} {2003})}\BibitemShut {NoStop}%
\bibitem [{\citenamefont {Rankin}\ \emph {et~al.}(2014)\citenamefont {Rankin},
  \citenamefont {Avitabile}, \citenamefont {Baladron}, \citenamefont {Faye},\
  and\ \citenamefont {Lloyd}}]{Rankin:2014bz}%
  \BibitemOpen
  \bibfield  {author} {\bibinfo {author} {\bibfnamefont {J.}~\bibnamefont
  {Rankin}}, \bibinfo {author} {\bibfnamefont {D.}~\bibnamefont {Avitabile}},
  \bibinfo {author} {\bibfnamefont {J.}~\bibnamefont {Baladron}}, \bibinfo
  {author} {\bibfnamefont {G.}~\bibnamefont {Faye}}, \ and\ \bibinfo {author}
  {\bibfnamefont {D.~J.~B.}\ \bibnamefont {Lloyd}},\ }\href@noop {} {\bibfield
  {journal} {\bibinfo  {journal} {SIAM J. Sci. Comput.}\ }\textbf {\bibinfo
  {volume} {36}},\ \bibinfo {pages} {B70} (\bibinfo {year} {2014})}\BibitemShut
  {NoStop}%
\bibitem [{\citenamefont {Bressloff}(2001)}]{Bressloff:2001ck}%
  \BibitemOpen
  \bibfield  {author} {\bibinfo {author} {\bibfnamefont {P.~C.}\ \bibnamefont
  {Bressloff}},\ }\href@noop {} {\bibfield  {journal} {\bibinfo  {journal}
  {Phys. D}\ }\textbf {\bibinfo {volume} {155}},\ \bibinfo {pages} {83}
  (\bibinfo {year} {2001})}\BibitemShut {NoStop}%
\bibitem [{\citenamefont {Coombes}\ and\ \citenamefont
  {Laing}(2011)}]{Coombes2011aa}%
  \BibitemOpen
  \bibfield  {author} {\bibinfo {author} {\bibfnamefont {S.}~\bibnamefont
  {Coombes}}\ and\ \bibinfo {author} {\bibfnamefont {C.}~\bibnamefont
  {Laing}},\ }\href@noop {} {\bibfield  {journal} {\bibinfo  {journal} {Phys.
  Rev. E}\ }\textbf {\bibinfo {volume} {83}},\ \bibinfo {pages} {011912}
  (\bibinfo {year} {2011})}\BibitemShut {NoStop}%
\bibitem [{\citenamefont {Avitabile}\ and\ \citenamefont
  {Schmidt}(2015)}]{Avitabile2015aa}%
  \BibitemOpen
  \bibfield  {author} {\bibinfo {author} {\bibfnamefont {D.}~\bibnamefont
  {Avitabile}}\ and\ \bibinfo {author} {\bibfnamefont {H.}~\bibnamefont
  {Schmidt}},\ }\href@noop {} {\bibfield  {journal} {\bibinfo  {journal} {Phys.
  D}\ }\textbf {\bibinfo {volume} {294}},\ \bibinfo {pages} {24} (\bibinfo
  {year} {2015})}\BibitemShut {NoStop}%
\bibitem [{\citenamefont {Laing}\ \emph {et~al.}(2002)\citenamefont {Laing},
  \citenamefont {Troy}, \citenamefont {Gutkin},\ and\ \citenamefont
  {Ermentrout}}]{Laing2002aa}%
  \BibitemOpen
  \bibfield  {author} {\bibinfo {author} {\bibfnamefont {C.~R.}\ \bibnamefont
  {Laing}}, \bibinfo {author} {\bibfnamefont {W.~C.}\ \bibnamefont {Troy}},
  \bibinfo {author} {\bibfnamefont {B.}~\bibnamefont {Gutkin}}, \ and\ \bibinfo
  {author} {\bibfnamefont {G.~B.}\ \bibnamefont {Ermentrout}},\ }\href@noop {}
  {\bibfield  {journal} {\bibinfo  {journal} {SIAM J. Appl. Math.}\ }\textbf
  {\bibinfo {volume} {63}},\ \bibinfo {pages} {62} (\bibinfo {year}
  {2002})}\BibitemShut {NoStop}%
\bibitem [{\citenamefont {Coombes}\ \emph {et~al.}(2003)\citenamefont
  {Coombes}, \citenamefont {Lord},\ and\ \citenamefont {Owen}}]{Coombes2003aa}%
  \BibitemOpen
  \bibfield  {author} {\bibinfo {author} {\bibfnamefont {S.}~\bibnamefont
  {Coombes}}, \bibinfo {author} {\bibfnamefont {G.}~\bibnamefont {Lord}}, \
  and\ \bibinfo {author} {\bibfnamefont {M.}~\bibnamefont {Owen}},\ }\href@noop
  {} {\bibfield  {journal} {\bibinfo  {journal} {Phys. D}\ }\textbf {\bibinfo
  {volume} {178}},\ \bibinfo {pages} {219} (\bibinfo {year}
  {2003})}\BibitemShut {NoStop}%
\bibitem [{\citenamefont {Coombes}\ \emph {et~al.}(2012)\citenamefont
  {Coombes}, \citenamefont {Schmidt},\ and\ \citenamefont
  {Bojak}}]{Coombes2012aa}%
  \BibitemOpen
  \bibfield  {author} {\bibinfo {author} {\bibfnamefont {S.}~\bibnamefont
  {Coombes}}, \bibinfo {author} {\bibfnamefont {H.}~\bibnamefont {Schmidt}}, \
  and\ \bibinfo {author} {\bibfnamefont {I.}~\bibnamefont {Bojak}},\
  }\href@noop {} {\bibfield  {journal} {\bibinfo  {journal} {J. Math.
  Neurosci.}\ }\textbf {\bibinfo {volume} {2}},\ \bibinfo {pages} {9} (\bibinfo
  {year} {2012})}\BibitemShut {NoStop}%
\bibitem [{\citenamefont {Amari}(1977)}]{Amari1977aa}%
  \BibitemOpen
  \bibfield  {author} {\bibinfo {author} {\bibfnamefont {S.-I.}\ \bibnamefont
  {Amari}},\ }\href@noop {} {\bibfield  {journal} {\bibinfo  {journal} {Biol.
  Cybern.}\ }\textbf {\bibinfo {volume} {27}},\ \bibinfo {pages} {77} (\bibinfo
  {year} {1977})}\BibitemShut {NoStop}%
\bibitem [{\citenamefont {Knobloch}(2015)}]{Knobloch15}%
  \BibitemOpen
  \bibfield  {author} {\bibinfo {author} {\bibfnamefont {E.}~\bibnamefont
  {Knobloch}},\ }\href@noop {} {\bibfield  {journal} {\bibinfo  {journal}
  {Annu. Rev. Condens. Matter Phys.}\ }\textbf {\bibinfo {volume} {6}},\
  \bibinfo {pages} {325} (\bibinfo {year} {2015})}\BibitemShut {NoStop}%
\bibitem [{\citenamefont {Brackley}\ and\ \citenamefont
  {Turner}(2007)}]{Brackley2007aa}%
  \BibitemOpen
  \bibfield  {author} {\bibinfo {author} {\bibfnamefont {C.~A.}\ \bibnamefont
  {Brackley}}\ and\ \bibinfo {author} {\bibfnamefont {M.~S.}\ \bibnamefont
  {Turner}},\ }\href@noop {} {\bibfield  {journal} {\bibinfo  {journal} {Phys.
  Rev. E}\ }\textbf {\bibinfo {volume} {75}},\ \bibinfo {pages} {{041913}}
  (\bibinfo {year} {2007})}\BibitemShut {NoStop}%
\bibitem [{\citenamefont {Thul}\ \emph {et~al.}(2016)\citenamefont {Thul},
  \citenamefont {Coombes},\ and\ \citenamefont {Laing}}]{Thul:2016gr}%
  \BibitemOpen
  \bibfield  {author} {\bibinfo {author} {\bibfnamefont {R.}~\bibnamefont
  {Thul}}, \bibinfo {author} {\bibfnamefont {S.}~\bibnamefont {Coombes}}, \
  and\ \bibinfo {author} {\bibfnamefont {C.~R.}\ \bibnamefont {Laing}},\
  }\href@noop {} {\bibfield  {journal} {\bibinfo  {journal} {J. Math.
  Neurosci.}\ }\textbf {\bibinfo {volume} {6}},\ \bibinfo {pages} {1} (\bibinfo
  {year} {2016})}\BibitemShut {NoStop}%
\bibitem [{\citenamefont {Coombes}\ and\ \citenamefont
  {Owen}(2005)}]{Coombes:2005hp}%
  \BibitemOpen
  \bibfield  {author} {\bibinfo {author} {\bibfnamefont {S.}~\bibnamefont
  {Coombes}}\ and\ \bibinfo {author} {\bibfnamefont {M.~R.}\ \bibnamefont
  {Owen}},\ }\href@noop {} {\bibfield  {journal} {\bibinfo  {journal} {Phys.
  Rev. Lett.}\ }\textbf {\bibinfo {volume} {94}},\ \bibinfo {pages} {148102}
  (\bibinfo {year} {2005})}\BibitemShut {NoStop}%
\bibitem [{\citenamefont {Coombes}\ and\ \citenamefont
  {Owen}(2007)}]{coombes2007exotic}%
  \BibitemOpen
  \bibfield  {author} {\bibinfo {author} {\bibfnamefont {S.}~\bibnamefont
  {Coombes}}\ and\ \bibinfo {author} {\bibfnamefont {M.~R.}\ \bibnamefont
  {Owen}},\ }in\ \href@noop {} {\emph {\bibinfo {booktitle} {Fluids and Waves:
  Recent Trends in Applied Analysis: Research Conference, May 11-13, 2006, the
  Universtiy of Memphis, Memphis, TN}}},\ Vol.\ \bibinfo {volume} {440}\
  (\bibinfo {organization} {American Mathematical Soc.},\ \bibinfo {year}
  {2007})\ p.\ \bibinfo {pages} {123}\BibitemShut {NoStop}%
\bibitem [{\citenamefont {Madison}\ and\ \citenamefont
  {Nicoll}(1984)}]{Madison:1984di}%
  \BibitemOpen
  \bibfield  {author} {\bibinfo {author} {\bibfnamefont {D.~V.}\ \bibnamefont
  {Madison}}\ and\ \bibinfo {author} {\bibfnamefont {R.~A.}\ \bibnamefont
  {Nicoll}},\ }\href@noop {} {\bibfield  {journal} {\bibinfo  {journal} {J.
  Physiol.}\ }\textbf {\bibinfo {volume} {354}},\ \bibinfo {pages} {319}
  (\bibinfo {year} {1984})}\BibitemShut {NoStop}%
\bibitem [{\citenamefont {Desroches}\ \emph {et~al.}(2016)\citenamefont
  {Desroches}, \citenamefont {Krupa},\ and\ \citenamefont
  {Rodrigues}}]{desroches2016spike}%
  \BibitemOpen
  \bibfield  {author} {\bibinfo {author} {\bibfnamefont {M.}~\bibnamefont
  {Desroches}}, \bibinfo {author} {\bibfnamefont {M.}~\bibnamefont {Krupa}}, \
  and\ \bibinfo {author} {\bibfnamefont {S.}~\bibnamefont {Rodrigues}},\
  }\href@noop {} {\bibfield  {journal} {\bibinfo  {journal} {Phys. D}\ }\textbf
  {\bibinfo {volume} {331}},\ \bibinfo {pages} {58} (\bibinfo {year}
  {2016})}\BibitemShut {NoStop}%
\bibitem [{\citenamefont {Avitabile}\ \emph {et~al.}(2017)\citenamefont
  {Avitabile}, \citenamefont {Nicks},\ and\ \citenamefont
  {Smith}}]{Avitabile:2016}%
  \BibitemOpen
  \bibfield  {author} {\bibinfo {author} {\bibfnamefont {D.}~\bibnamefont
  {Avitabile}}, \bibinfo {author} {\bibfnamefont {R.}~\bibnamefont {Nicks}}, \
  and\ \bibinfo {author} {\bibfnamefont {O.}~\bibnamefont {Smith}},\
  }\href@noop {} {\bibfield  {journal} {\bibinfo  {journal} {in preparation}\ }
  (\bibinfo {year} {2017})}\BibitemShut {NoStop}%
\bibitem [{\citenamefont {Coombes}\ \emph {et~al.}(2013)\citenamefont
  {Coombes}, \citenamefont {Schmidt},\ and\ \citenamefont
  {Avitabile}}]{Coombes2013aa}%
  \BibitemOpen
  \bibfield  {author} {\bibinfo {author} {\bibfnamefont {S.}~\bibnamefont
  {Coombes}}, \bibinfo {author} {\bibfnamefont {H.}~\bibnamefont {Schmidt}}, \
  and\ \bibinfo {author} {\bibfnamefont {D.}~\bibnamefont {Avitabile}},\ }in\
  \href@noop {} {\emph {\bibinfo {booktitle} {Neural Field Theory}}},\ \bibinfo
  {editor} {edited by\ \bibinfo {editor} {\bibfnamefont {S.}~\bibnamefont
  {Coombes}}, \bibinfo {editor} {\bibfnamefont {P.}~\bibnamefont {beim
  Graben}}, \bibinfo {editor} {\bibfnamefont {R.}~\bibnamefont {Potthast}}, \
  and\ \bibinfo {editor} {\bibfnamefont {J.~J.}\ \bibnamefont {Wright}}}\
  (\bibinfo  {publisher} {Springer},\ \bibinfo {address} {New York},\ \bibinfo
  {year} {2013})\ pp.\ \bibinfo {pages} {187--211}\BibitemShut {NoStop}%
\bibitem [{\citenamefont {Folias}\ and\ \citenamefont
  {Bressloff}(2005{\natexlab{b}})}]{Folias2005aaa}%
  \BibitemOpen
  \bibfield  {author} {\bibinfo {author} {\bibfnamefont {S.~E.}\ \bibnamefont
  {Folias}}\ and\ \bibinfo {author} {\bibfnamefont {P.~C.}\ \bibnamefont
  {Bressloff}},\ }\href@noop {} {\bibfield  {journal} {\bibinfo  {journal}
  {SIAM J. Appl. Math.}\ }\textbf {\bibinfo {volume} {65}},\ \bibinfo {pages}
  {2067} (\bibinfo {year} {2005}{\natexlab{b}})}\BibitemShut {NoStop}%
\bibitem [{\citenamefont {Haimovici}\ \emph {et~al.}(2013)\citenamefont
  {Haimovici}, \citenamefont {Tagliazucchi}, \citenamefont {Balenzuela},\ and\
  \citenamefont {Chialvo}}]{Haimovici2013}%
  \BibitemOpen
  \bibfield  {author} {\bibinfo {author} {\bibfnamefont {A.}~\bibnamefont
  {Haimovici}}, \bibinfo {author} {\bibfnamefont {E.}~\bibnamefont
  {Tagliazucchi}}, \bibinfo {author} {\bibfnamefont {P.}~\bibnamefont
  {Balenzuela}}, \ and\ \bibinfo {author} {\bibfnamefont {D.~R.}\ \bibnamefont
  {Chialvo}},\ }\href@noop {} {\bibfield  {journal} {\bibinfo  {journal} {Phys.
  Rev. Lett.}\ }\textbf {\bibinfo {volume} {110}},\ \bibinfo {pages} {178101}
  (\bibinfo {year} {2013})}\BibitemShut {NoStop}%
\bibitem [{\citenamefont {Proix}\ \emph {et~al.}(2014)\citenamefont {Proix},
  \citenamefont {Bartolomei}, \citenamefont {Chauvel}, \citenamefont
  {Bernard},\ and\ \citenamefont {Jirsa}}]{proix2014}%
  \BibitemOpen
  \bibfield  {author} {\bibinfo {author} {\bibfnamefont {T.}~\bibnamefont
  {Proix}}, \bibinfo {author} {\bibfnamefont {F.}~\bibnamefont {Bartolomei}},
  \bibinfo {author} {\bibfnamefont {P.}~\bibnamefont {Chauvel}}, \bibinfo
  {author} {\bibfnamefont {C.}~\bibnamefont {Bernard}}, \ and\ \bibinfo
  {author} {\bibfnamefont {V.~K.}\ \bibnamefont {Jirsa}},\ }\href@noop {}
  {\bibfield  {journal} {\bibinfo  {journal} {J. Neurosci.}\ }\textbf {\bibinfo
  {volume} {34}},\ \bibinfo {pages} {15009} (\bibinfo {year}
  {2014})}\BibitemShut {NoStop}%
\bibitem [{\citenamefont {Borst}\ and\ \citenamefont
  {Theunissen}(1999)}]{borst1999}%
  \BibitemOpen
  \bibfield  {author} {\bibinfo {author} {\bibfnamefont {A.}~\bibnamefont
  {Borst}}\ and\ \bibinfo {author} {\bibfnamefont {F.~E.}\ \bibnamefont
  {Theunissen}},\ }\href@noop {} {\bibfield  {journal} {\bibinfo  {journal}
  {Nat. Neurosci.}\ }\textbf {\bibinfo {volume} {2}},\ \bibinfo {pages} {947}
  (\bibinfo {year} {1999})}\BibitemShut {NoStop}%
\end{thebibliography}%

\end{document}